\documentclass[useAMS,usenatbib]{mn2e}
\usepackage{epsfig}
\usepackage{amssymb}
\usepackage{amsmath}
\usepackage{graphicx,subfigure}
\usepackage{caption}
\usepackage{tabto}
\usepackage{tabularx}
\usepackage{longtable}
\usepackage{lscape}
\usepackage{booktabs}
\usepackage{titlesec}
\usepackage{float}
\usepackage{url}
\usepackage{color}
\def\Msun{\ifmmode{{\rm M}_\odot}\else${\rm M}_\odot$\fi}
\def\msun{\ifmmode{{\rm M}_\odot}\else${\rm M}_\odot$\fi}

\title[Genesis of magnetic fields]
{Genesis of magnetic fields in isolated white dwarfs}

\author[Briggs, Ferrario, Tout \& Wickramasinghe] {Gordon
  P.~Briggs$^1$, Lilia Ferrario$^1$, Christopher A.~Tout$^{1,2,3}$,
  \newauthor Dayal T.~Wickramasinghe$^1$\\
  $^1$Mathematical Sciences Institute, The Australian National
  University, ACT 0200, Australia\\
  $^2$Institute of Astronomy, The Observatories, Madingley Road, Cambridge CB3 0HA\\
  $^3$Monash Centre for Astrophysics, School of Physics and Astronomy,
  10 College Walk, Monash University 3800, Australia}

\begin{document}

\date{Accepted.  Received ; in original form} 

\maketitle

\label{firstpage}

\begin{abstract}
  A dynamo mechanism driven by differential rotation when stars merge
  has been proposed to explain the presence of strong fields in
  certain classes of magnetic stars. In the case of the high field
  magnetic white dwarfs (HFMWDs), the site of the differential
  rotation has been variously thought to be the common envelope, the
  hot outer regions of a merged degenerate core or an accretion disc
  formed by a tidally disrupted companion that is subsequently
  accreted by a degenerate core. We have shown previously that the
  observed incidence of magnetism and the mass distribution in HFMWDs
  are consistent with the hypothesis that they are the result of
  merging binaries during common envelope evolution. Here we calculate
  the magnetic field strengths generated by common envelope
  interactions for synthetic populations using a simple prescription
  for the generation of fields and find that the observed magnetic
  field distribution is also consistent with the stellar merging
  hypothesis. We use the Kolmogorov-Smirnov test to study the
  correlation between the calculated and the observed field strengths
  and find that it is consistent for low envelope ejection
  efficiency. We also suggest that field generation by the plunging of
  a giant gaseous planet on to a white dwarf may explain why magnetism
  among cool white dwarfs (including DZ\,white dwarfs) is higher than
  among hot white dwarfs. In this picture a super-Jupiter residing in
  the outer regions of the white dwarf's planetary system is perturbed
  into a highly eccentric orbit by a close stellar encounter and is
  later accreted by the white dwarf.

\end{abstract}

\section{Introduction}
\label{Introd}

The existence of strong magnetic fields in stars at any phase of
their evolution is still largely unexplained and very puzzling
\citep[see][]{Ferrario2015a,Wick2000}.  High field magnetic white
dwarfs (HFMWDs) have dipolar magnetic field strengths of up to
$10^9$\,G.  There are no observed HFMWDs with late-type
companions found in wide binary systems.
\citet{Liebert2005,Liebert2015} pointed out that this contrasts with
non-magnetic white dwarfs, a large fraction of which are found in such
systems.  This led \citet{Tout2008} to hypothesise that the entire
class of HFMWDs with fields $10^6 < B/{\rm G} < 10^ 9$ owe their
magnetic fields to binary systems which have merged while in a common
envelope stage of evolution. In this scenario, when one of the two
stars in a binary evolves to become a giant or a super-giant its
expanded outer layers fill its Roche lobe.  At this point unstable
mass transfer leads to a state in which the giant's envelope engulfs
the companion star as well as its own core. This merging idea to
explain the origin of fields in white dwarfs is now favoured over
the fossil field hypothesis first suggested by \citet{ang1981} whereby
the magnetic main-sequence Ap and Bp stars are the ancestors of the
HFMWDs if magnetic flux is conserved all the way to the compact star
phase \citep[see also][and references
therein]{Tout2004,Wickramasinghe2005}.
  
During common envelope evolution, frictional drag forces acting on the
cores and the envelope cause the orbit to decay. The two cores spiral
together losing energy and angular momentum which are transferred to
the differentially revolving common envelope, part of which at least,
is ejected from the system.  This process is thought to proceed on a
dynamical time scale of less than a few thousand years and hence has
never been observed.  The original model of \citet{Tout2008} suggested
that high fields were generated by a dynamo between the common
envelope and the outer layers of the proto-white dwarf before the
common envelope is ejected. If the cores merge the resulting giant
star eventually loses its envelope to reveal a single HFMWD. If the
envelope is ejected when the cores are close but have not merged a
magnetic CV is formed. \citet{potter2010} found problems with
  this scenario in that the time-scale for diffusion of the field into
  the white dwarf is significantly longer than the expected common
  envelope lifetime.  Instead \citet{WTF2014} suggested that a weak
  seed field is intensified by the action of a dynamo arising from the
  differential rotation in the merged object as it forms.  This dynamo
  predicts a poloidal magnetic flux that depends only on the initial
  differential rotation and is independent of the initial
  field. \citet{nordhaus2011} suggested another model where magnetic
  fields generated in an accretion disc formed from a tidally
  disrupted low-mass companion are advected on to the surface of the
  proto-white dwarf. However, this would once again depend on the
  time-scale for diffusion of the field into the surface layers of the
  white dwarf.  \citet{garcia2012} found that a field of about
$3\times10^{10}$\,G could be created from a massive, hot and
differentially rotating corona forming around a merged DD. They also
carried out a population synthesis study of merging DDs with a common
envelope efficiency factor $\alpha=0.25$.  They achieved good
agreement in the observed properties between high-mass white dwarfs
($M_\textrm{WD}\ge$ 0.8\msun) and HFMWDs but their studies did not
include degenerate cores merging with non-degenerate companions as did
\citet[][hereinafter paper\,I]{Briggs2015}.

The stellar merging hypothesis may only apply to
HFMWDs. \citet{landstreet2012} point out that weak fields of
$B\le1$\,kG may exist in most white dwarfs and so probably arise in
the course of normal stellar evolution from a dynamo action between
the core and envelope.

With population synthesis we showed, in paper\,I, that the origin of
HFMWDs is consistent with the stellar merging hypothesis. The
calculations presented in paper\,I could explain the observed incidence of
magnetism among white dwarfs and showed that the computed mass
distribution fits the observed mass distribution of the HFMWDs
more closely than it fits the mass distribution of non-magnetic white
dwarfs. This demonstrated that magnetic and non-magnetic white dwarfs
belong to two populations with different progenitors. We now present
the results of calculations of the magnetic field strength expected
from merging binary star systems.

\section{Population synthesis calculations}
\label{sec:calculations}

As described in paper\,I, we create a population of binary systems by
evolving them from the zero-age main sequence (ZAMS) to 9.5\,Gyr, the
age of the Galactic disc \citep{Kilic2017}. Often an age of 12\,Gyr
is assumed when population synthesis studies are carried out but an
integration age of 12\,Gyr, that encompasses not only the thin and
thick disc but also the inner halo, would be far too large for our
studies of the origin of HFMWDs. The HFMWDs belong to the thin disc population,
according to the kinematic studies of HFMWDs by \citet{Sion1988} and
\citet{Anselowitz1999},  who found that HFMWDs come from a young
stellar disc population characterised by small motions with respect to
the Sun and a dearth of genuine old disc and halo space
velocities. The more recent studies of the white dwarfs within 20\,pc
of the Sun by \citet{Sion2009} also support the earlier findings and
show that the HFMWDs in the local sample have significantly lower
space velocities than non-magnetic white dwarfs.

We use the rapid binary stellar evolution algorithm {\sc bse}
developed by \citet{Hurley2002} that allows modelling of the most
intricate binary evolution.  This algorithm includes not only
all those features that characterise the evolution of single stars
\citep{Hurley2000} but also all major phenomena pertinent to binary
evolution. These comprise Roche lobe overflow, common envelope
evolution \citep{Paczynski1976}, tidal interaction, collisions,
gravitational radiation and magnetic braking.

As in paper\,I, we have three initial parameters. The mass of the
primary star $0.8\le M_1/\msun\le 12.0$, the mass of its companion
$0.1\le M_2/\msun\le 12.0$ and the orbital period
$0.1\le P_0/{\rm d}\le 10\,000$. These initial parameters are on a
logarithmic scale of 200\,divisions.  We then compute the real number
of binaries assuming that the initial mass of the primary star is
distributed according to Salpeter's (1955) mass function and the
companion's mass according to a flat mass ratio distribution with
$q\le 1$ \citep[e.g.][]{Hurley2002,Ferrario2012}. The period
distribution is taken to be uniform in its logarithm. We use the
efficiency parameter $\alpha$ (energy) formalism for the common
envelope phases with $\alpha$ taken as a free parameter between $0.1$
and $0.9$.  In our calculations we have used $\eta=1.0$ for the
Reimers' mass-loss parameter and a stellar metallicity $Z = 0.02$.  We
select a sub-population consisting of single white dwarfs that formed
by merging during common envelope evolution. Conditions of the
selection are that (i) at the beginning of common envelope evolution
the primary has a degenerate core to ensure that any magnetic field
formed or amplified during common envelope persists in a frozen-in
state and (ii) from the end of common envelope to the final white
dwarf stage there is no further nuclear burning in the core of the
pre-white dwarf star which would otherwise induce convection that
would destroy any frozen-in magnetic field. In addition to stellar
merging during common envelope, we also select double white dwarf
binaries whose components merge to form a single white dwarf at any
time after the last common envelope evolution up to the age of the
Galactic disc. This forms our DD merging channel for the formation of
HFMWDs.

\subsection{Theoretical magnetic field strength}\label{TheoryB}

\begin{figure}
\centering
\includegraphics[width=1.00\columnwidth]{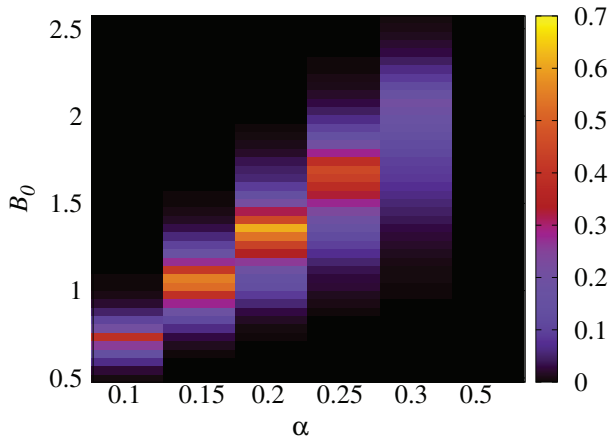}
\caption{Density plot of the probability given by the K--S test that
  the CDFs of the theoretical and observed magnetic field
  distributions are drawn from the same population. This was generated
  for a range of $\alpha$ and $B_0$ (see text). The probability is
  colour-coded according to the palette shown on the right hand side
  of the figure. The sub-structures in this plot are caused by the
  discretisation of $\alpha$ and $B_0$.}
\label{fig:contour}
\end{figure}

The goal of this paper is to construct the magnetic field distribution
of our synthetic sample of HFMWDs using, as a basis, the results and
ideas set out by \citet{Tout2008} and Wickramasinghe et al. (2014) .  If the cores
of the two stars do not merge during common envelope, our assumption
is that a fraction of the maximum angular momentum available at the
point of the ejection of the envelope causes the shear necessary to
generate the magnetic field. The non-merging case, leading to the
formation of MCVs, is presented by \citet[][
paper\,III]{Briggs2018}. In the case of coalescing cores, a fraction
of the break-up angular momentum of the resulting degenerate core
provides the shear required to give rise to the strongest fields. In
the following sections and in paper\,III we show that our models
indeed show that the highest fields are generated when two stars merge
and give rise to a HFMWD.

Having obtained the actual number of white dwarfs we then assign a
magnetic field $B$ to each. Our prescription is that the field,
generated and acquired by the white dwarf during common envelope
evolution or DD merging, is proportional to the orbital angular
velocity $\Omega=\displaystyle{\frac{2\pi}{P_{\rm orb}}}$ of the
binary at the point the envelope is ejected and write
\begin{equation}\label{EqBfield}
B = B_0\left(\frac{\Omega}{\Omega_{\rm crit}}\right)\, \mbox{G}.
\end{equation}
where 
\begin{equation}\label{omega_c}
	\Omega_{\rm crit}= \sqrt{\frac{GM_{\rm WD}}{R_{\rm
              WD}^3}}=0.9\left(\frac{M_{\rm
              WD}}{\Msun}\right)^{1/2}\left(\frac{5.4\times 10^8}{R_{\rm WD}}\right)^{-3/2}
\end{equation}
is the break-up angular velocity of a white dwarf of mass $M_{\rm WD}$
and radius $R_{\rm WD}$.

This model encapsulates the dynamo model of Wickramasinghe et
  al. (2014) where a seed poloidal field is amplified to a maximum
  that depends \emph{linearly} on the initial differential rotation
  imparted to the white dwarf. In view of these results, here we
  simply assume a linear relationship between the poloidal field and
  the initial rotation and recalibrate the Wickramasinghe's et
  al. (2014) relation between differential rotation and field using
  (i) a more recent set of data and (ii) results from our population
  synthesis calculations that provide $\Omega$ in equation
  (\ref{EqBfield}). The quantity $B_0$ in equation (\ref{EqBfield}) is
  also a parameter to be determined empirically.  Different $B_0$'s
  simply shift the field distribution to lower or higher fields with
  no changes to the shape of the field distribution which is solely
  determined by the common envelope efficiency parameter $\alpha$.

For the radius of the white dwarf we use Nauenberg's (1972)
mass-radius formula
\begin{equation}\label{R_WD}
R_{\rm WD}= 0.0112 R_\odot\,\left[\left(\frac{M_{\rm Ch}}{M_{\rm WD}}\right)^{2/3}-\,\,\,\left(\frac{M_{\rm WD}}{M_{\rm Ch}}\right)^{2/3}\right]^{1/2},
\end{equation}
where $M _{\rm{Ch}}=1.44$\,M$_\odot$ is the Chandrasekhar limiting
mass.

\begin{figure}
\centering
\includegraphics[width=1.00\columnwidth]{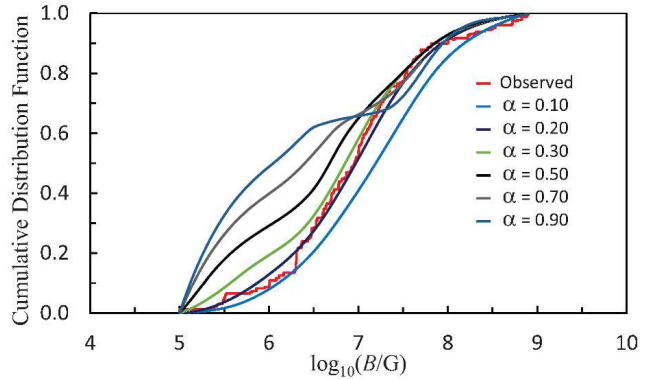}
\caption{CDFs of observed (red) and {\sc{BSE}} theoretical magnetic field
  distributions for a Galactic disc age of 9.5 Gyr and various
  $\alpha$.}
\label{fig:CDF_Magfield}
\end{figure}

\subsection{Parameters calibration}\label{Calibration}

The data set of HFMWDs is affected by many biases, even though
  some of the surveys that discovered them were
  magnitude-limited. This is because HFMWDs tend to be more massive
  than their non-magnetic counterparts, as first noticed by
  \citet{Liebert1988}, and therefore their smaller radii, as expected
  by equation (\ref{R_WD}), make them dimmer and so less likely to be
  detected.  Volume-limited samples are far better, given that our
  synthetic population mimics a volume-limited sample, but do not
  include enough HFMWDs to allow us to conduct any statistically
  meaningful study.  In this section we establish the parameter space
  of relevance to the observations of HFMWDs by comparing the
  predictions of the magnetic field distribution derived from our
  population synthesis calculations to the fields of HFMWDs listed in
  \citet{Ferrario2015b}. In order to achieve this goal we have
employed the Kolmogorov--Smirnov (K--S) test \citep{Press1992} to
establish which combination of $B_0$ and $\alpha$ yield the best fit
to the observed field distribution of HFMWDs.  The K--S test compares
the cumulative distribution functions (CDFs) of two data samples (in
this case the theoretical and observed field distributions) and gives
the probability $P$ that they are drawn randomly from the same
population. We have calculated CDFs for seven different $\alpha$ and
44 different $B_0$s for each $\alpha$. If we discard all combinations
of $\alpha$ and $B_0$ for which $P\le0.01$, we find
$0.5 \times 10^{10} \le B_0/{\rm G} \le 2.5 \times10^{10}$ and
$\alpha<0.5$. We have depicted in Fig.\,\ref{fig:contour} a density
plot of our results.  The highest probability is for
$B_0=1.35\times10^{10}$\,G and $\alpha=0.2$.  We show in
Fig.\,\ref{fig:CDF_Magfield} the theoretical CDFs for
$B_0=1.35\times10^{10}$\,G and various $\alpha$s and the CDF of the
observations of the magnetic field strengths of HFMWDs.

In the following sections we will discuss models with
  $B_0=1.35\times10^{10}$\,G and a range of $\alpha$ again noting that
  a different $B_0$ would simply move the field distribution to
  lower or higher fields with no change of shape. Therefore our
  discussion in the following sections will focus on the effects of
  varying $\alpha$.

\begin{figure*}
\centering
\includegraphics[width=0.95\textwidth]{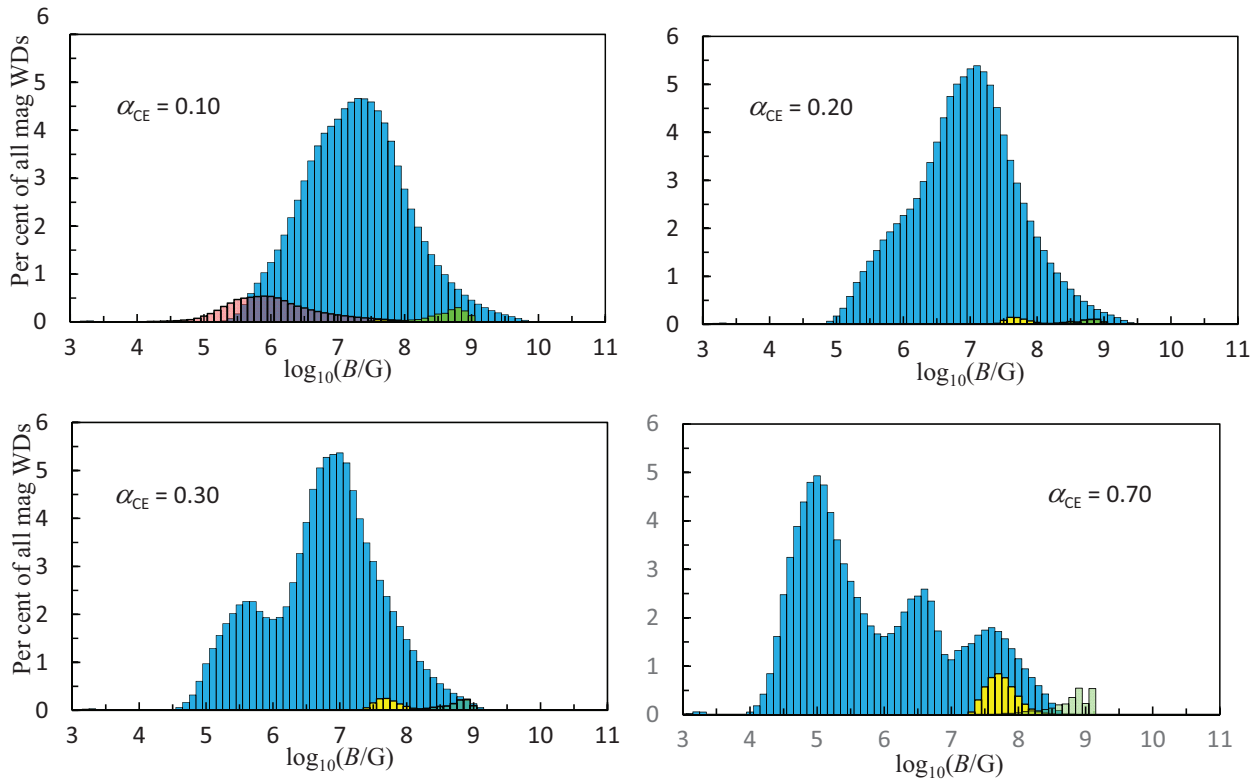}
\caption {Theoretical magnetic field strength for a Galactic disc age
  of 9.5 Gyr and various $\alpha$. The histograms are superimposed,
  not stacked, to highlight the contribution made by each type of
  white dwarf to the overall distribution. The blue, red and yellow
  histograms represent, respectively, CO, ONe, He white dwarfs. The
  green histograms depict the merged DD systems.}
\label{fig:MagAlphas}
\end{figure*}

\section{Discussion of results}\label{MagFieldDistr}

Fig. \ref{fig:MagAlphas} shows the calculated magnetic field
distribution and the breakdown of the WD types for $\alpha=0.1$ to
$0.7$. The maximum field strength is a few $10^{9}$\,G and is found
mostly in systems in which the HFMWD forms either via the merging of
two very compact stars on a tight orbit or through the merging of two
white dwarfs after common envelope evolution (DD path). The reason for
this is that these systems have very short periods and when they merge
produce very strongly magnetic WDs, as expected from equation
(\ref{EqBfield}).

\begin{figure*}
\centering
\includegraphics[width=0.95\textwidth]{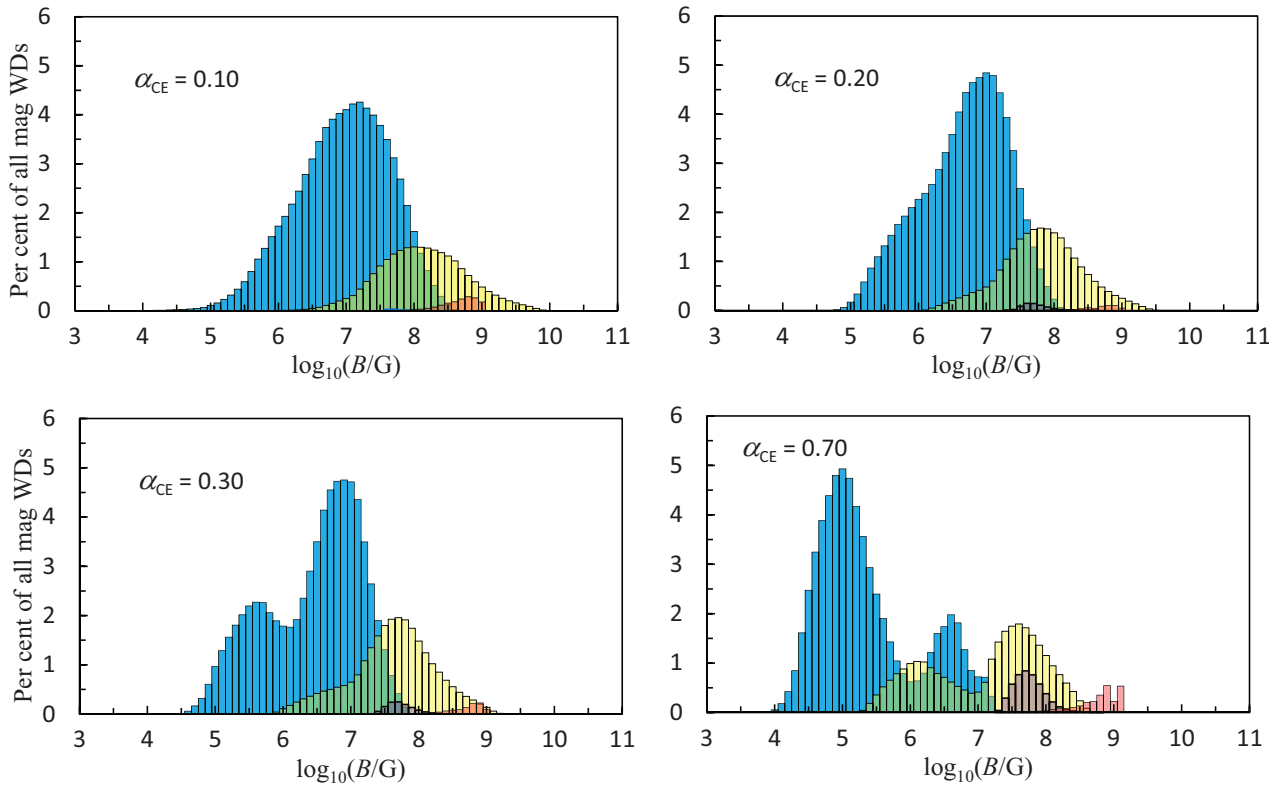}
\caption{Theoretical magnetic field distribution of HFMWDs showing the
  pre-common envelope progenitors for various $\alpha$. The light blue,
  yellow and purple histograms represent, respectively, the AGB/MS, AGB/CS and RGB/CS
  merging pairs. The red histograms depict the merged DD systems.}
\label{fig:paths}
\end{figure*}

We show in Fig.\,\ref{fig:paths} the theoretical magnetic field
distribution of HFMWDs for $\alpha=0.1$ to $\alpha=0.7$ with the
breakdown of their main formation channels, that is, their pre-common
envelope progenitors. The overwhelming contributors to the HFMWD
population are asymptotic giant branch (AGB) stars merging with
main-sequence (MS) or deeply convective stars (CS). At low $\alpha$,
systems with initially short orbital periods merge as soon as their
primaries evolve off the main sequence, either whilst in the
Hetzsprung's gap or during their ascent along the red giant branch
(RGB). Usually such merging events produce single stars that continue
their evolution burning helium in their cores and later on, depending
on the total mass of the merged star, heavier elements. Because of
core nuclear burning these stars continue their evolution to
eventually become single non-magnetic white dwarfs. The only
observational characteristic that may distinguish them from other
non-magnetic white dwarfs could be an unusual mass that does not fit
any reasonable initial to final mass function associated to the
stellar cluster to which they belong. On the other hand, if the RGB
star has a degenerate core, as for stars with $M_1\le 2.2$\,M$_\odot$
on the ZAMS, and merges with a low-mass CS, then the resulting object
is a strongly magnetic He\,WD. These RGB/CS merging events do
occur at all $\alpha$ but their fraction is higher at large $\alpha$
owing to fewer overall merging occurrences at high envelope clearance
efficiencies.

When systems do not merge when the primary evolves on the RGB, they
may merge when they undergo common envelope evolution on the AGB. In
this case those binaries with the shortest orbital periods at the
beginning of the common envelope evolution are those that form the
highest magnetic field tail of the distribution. There are two main
types of merging pairs, AGB stars merging with MS stars
($M\ge0.7$\,M$_\odot$) and AGB stars merging with CS
($M<0.7$\,M$_\odot$). Each of these combinations exhibits two peaks as
seen in Fig.\,\ref{fig:paths} for $\alpha>0.2$, although the second peak at
lower fields of the merging AGB/CS pair becomes well defined only when
$\alpha=0.7$.  Because AGB/MS systems have larger orbital periods
at the onset of common envelope evolution, their merging gives rise to
generally more massive but less magnetic white dwarfs as expected from
equation (\ref{EqBfield}). This is why the bulk of AGB/MS merging
pairs occupy the lowest and most prominent peak near $B=10^{5.5}$\,G
with the secondary maximum at $B=10^{6.8}$\,G. The AGB/CS merging
pairs form another two peaks, one at $B=10^6$\,G and the other at
$B=10^{7.75}$\,G. RGB stars merging with CS stars also form a maximum
at $B=10^{7.75}$\,G.  The reason for the double peaks in AGB/MS and
AGB/CS merging pairs is because high envelope clearance efficiencies
(high $\alpha$) require more massive primaries to bring the two stars
close enough together to merge during common envelope evolution. Thus,
these double peaks are caused by a dearth of AGB/MS merging pairs near
$B=10^6$\,G and of AGB/CS pairs near $B=10^7$\,G. Those systems whose
orbital periods would give rise to magnetic fields in these gaps fail
to merge because their initial periods are large and their primary
stars are not massive enough to bring the two components close enough
to merge. These double peaks are not present at low $\alpha$ 
because low envelope clearance efficiency always leads to tighter
orbits and merging is more likely for a much wider range of
initial masses and orbital periods, more effectively smearing the
contributions made by specific merging pairs.

\section{Comparison to observations}\label{Comparison}

A prediction of our merging hypothesis for the origin of HFMWDs is
that low-mass HFMWDs, mostly arising from AGB/CS merging pairs, should
display fields on average stronger than those of massive HFMWDs which
predominantly result from the merging of AGB/MS pairs. The HFMWDs
formed through the merging of two white dwarfs (DD channel) are
excluded from this prediction.  These are expected to produce objects
that are on average more massive, more strongly magnetic, and may be
spinning much faster than most HFMWDs
\citep[e.g. RE\,J0317-853,][]{Barstow1995,Ferrario1997,Vennes2003}. Given
the very small number of HFMWDs for which both mass and field are
known, it is not possible to verify whether this trend is present in
observed in HFMWDs. The problem is that it is very difficult to
measure masses of HFMWDs when their field is above a few $10^6$\,G. In
the low field regime one can assume that each Zeeman component is
broadened as in the zero field case. That is, the field does not
influence the structure of the white dwarf's atmosphere. Thus, the
modelling of Zeeman spectra has allowed us the determination of masses
and temperatures of lower field white dwarfs such as
1RXS\,J0823.6−2525 \citep[$B=2.8-3.5$\,MG and
M=1.2\,\msun;][]{Ferrario1998}, PG\,1658+441 \citep[$B=3.5$\,MG and
M=1.31\,\msun;][]{Schmidt1992,Ferrario1998} and the magnetic component
of the double degenerate system NLTT\,12758 \citep[$B=3.1$\,MG and
$M=0.69$\,\msun;][]{Kawka2017}. The masses of high field objects can
only be determined when their trigonometric parallax is known
\citep[e.g. Grw\,+70$^\circ$8247 with $B=320\pm20$\,MG and
$M=0.95\pm0.02$\,M$_\odot$,][]{Greenstein1985,Wickramasinghe1988}.
Nevertheless, it is encouraging to see that all the most massive (near
the Chandrasekhar's limit) currently known HFMWDs do indeed possess
low field strengths and that the merged DD RE\,J0317-853 is a strongly
magnetic white dwarf. A test of our prediction of an inverse relation
between field strength and mass will become possible with the release
of the accurate astrometric data of a billion stars by the ESA
satellite {\it Gaia}. This new set of high quality data will not only allow
us to test the (non-magnetic) white dwarf mass--radius relation but
will also provide us with precise mass and luminosity measurements of
most of the currently known white dwarfs, including the HFMWDs
\citep{Jordan2007}.

\begin{figure}
\centering
\includegraphics[width=1.00\columnwidth]{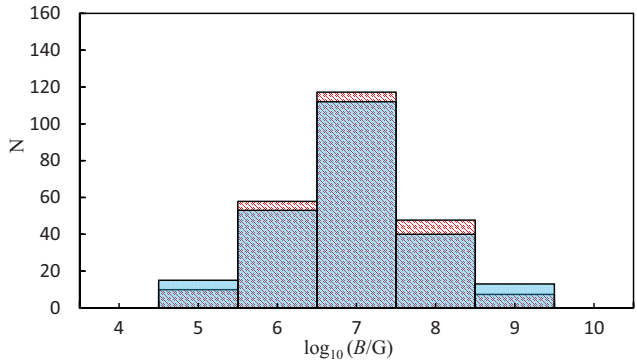}
\caption{Theoretical field distribution for $\alpha = 0.2$ of the
  total of the four types of HFMWDs (pink histogram) compared to the
  field distribution of the observed HFMWDs (blue histogram). }
  \label{fig:Theory_OBS_HFMWD} 
\end{figure}

The theoretical distribution for $\alpha=0.2$ overlapped to the
observations of HFMWDs is displayed in
Fig.\,\ref{fig:Theory_OBS_HFMWD}.  This figure shows that the maxima
of the theoretical and observed distributions occur near the same
field strength with the theoretical distribution extending from
$10^5$\,G to $10^9$\,G, as observed. The overwhelming contribution to
the theoretical field distribution is from CO\,WDs (see
Fig.\,\ref{fig:MagAlphas}).  ONe\,WDs are the next most common but at
much lower frequency and with field strengths
$4\le\log_{10} B/{\rm G} \le 8$.  Merged DD white dwarfs present field
strengths $8\le\log_{10}B/{\rm G} \le 9$ at an even lower frequency
than the ONe\,WDs.  Finally, He\,WDs are present in very small numbers
with field strengths centred at $B=10^{7.75}$\,G. This is in contrast to
observations of HFMWDs that show the presence of very low-mass objects
\citep[see table\,1 of][]{Ferrario2015b} that the {\sc bse} formalism
is unable to form. This mismatch between theory and observations may
be corrected through the use of, e.g., different superwind
assumptions \citep[see][and references therein]{Han1994,Meng2008}.

We note that the models shown in Fig.\,\ref{fig:MagAlphas} with
  $\alpha>0.2$ predict the existence of a large fraction of
  low-field magnetic white dwarfs with a bump appearing near
  $B=10^{5.5}$\,G for $\alpha=0.3$. This bump shifts toward lower
  fields and becomes increasingly more prominent as $\alpha$
  increases. For $\alpha= 0.7$ this low-field hump is the most
  prominent feature of the magnetic field distribution. In the past
  suggestions were made that the incidence of magnetism in white
  dwarfs may be bimodal, sharply rising below $10^5$\,G with an
  incidence that was predicted to be similar to or exceeding that of
  HFMWDs \citep{Wick2000}. However, recent low-field
  spectropolarimetric surveys of white dwarfs have not found anywhere
  near the number of objects that had been forecast to exist in this
  low-field regime \citep{landstreet2012}. Therefore, there is enough
  observational evidence to allow us to exclude the bimodality of the
  magnetic field distribution that is theoretically predicted for
  large $\alpha$'s.

\section{Incidence of magnetism among cool white dwarfs}

Because white dwarfs have very high gravities, all chemical elements
heavier than hydrogen, helium and dredged-up carbon or oxygen, quickly
sink to the bottom of their atmosphere. Nonetheless, up to 30\,per
cent of white dwarfs exhibit traces of Ca, Si, Mg, Fe, Na and other
metals \citep[DZ\,white dwarfs,][]{Zuckerman2003}.  This metal
pollution has been attributed to the steady accretion of debris from
the tidal disruption of large asteroids and rocky planets
\citep{Jura2003} making these white dwarfs important tools for the
study of the chemical composition of exosolar planets.  Interestingly,
the incidence of magnetism among cool ($T_{\rm eff}<8\,000$\,K)
DZ\,white dwarfs is about 13\,per cent \citep{Kawka2014, Hollands2015}
which is much higher than between 2 and 5\,per cent in the general
white dwarf population \citep{Ferrario2015a}. Although our
  modelling does not include the merging of sub-stellar companions, we
  speculate that the moderately strong magnetic fields observed in
  metal-polluted white dwarfs
  \citep[$0.5\le B/10^7{\rm G}\le 1.1$,][]{Hollands2017} may be caused
  by giant gaseous planets plunging into the star. The accretion of
  other minor rocky bodies would then produce the observed atmospheric
  pollution.  This mechanism could be applicable to all white dwarfs,
  although it is not clear what the fraction of HFMWDs that may have
  undergone this process is. Currently only 10 \citep{Hollands2017}
  out of about 240 HFMWDs are metal-polluted. Such merging events may
occur during the latest stages of AGB evolution when the outer
envelope of the star engulfs the innermost planets and the drag forces
exerted on them as they move through the stellar envelope cause them
to drift toward the degenerate stellar core \citep{Li1998}. Whilst
this mechanism is plausible, it does not explain why the incidence of
magnetism is much higher among \emph{cool} DZ\,white dwarfs.  Another
possibility involves close stellar encounters able to significantly
disturb the orbits of outer planets and asteroid belts. Such
encounters can trigger dynamical instabilities that cause the inward
migration, and accretion by the white dwarf, of a massive gaseous
planet and other rocky planets and asteroids. Because it takes
hydrogen-rich white dwarfs with $0.5\le M/\Msun\le 1.0$ about
$1.5-9$\,billion years to reach effective temperatures between 5\,000
and 8\,000\,K \citep{Tremblay2011,Kowalski2006}, such stellar
encounters are possible, as discussed in detail by \citet{Farihi2011}
to explain the origin of the very cool ($T_{\rm eff}=5310$\,K) and
polluted magnetic white dwarf G77--50.

A similar explanation may be invoked to explain the high incidence of
magnetism among cool white dwarfs of all types, as first reported by
\citet{Liebert1979}. The study of \citet{Fabrika1999} showed that
whilst the incidence of magnetism among hot white dwarfs is only
around 3.5\,per cent, it increases above 20\,per cent among cool white
dwarfs. The volume-limited sample of \citet{Kawka2007} also shows a
high incidence of magnetism (greater than 10\,per cent) which is
consistent with the fact that volume-limited samples are dominated by
cooler objects.  Even the Palomar-Green magnitude-limited sample study
of \citet{Liebert2003} shows a higher incidence of magnetism among
cooler white dwarfs than hotter ones. Over the years this topic has
been a cause of concern.  It is difficult to think of how fields could
be generated once the star has already evolved into a white dwarf
because, if anything, fields decay over time.  Alternatively, one
could argue that the formation rate of HFMWDs was higher when the
Galactic disc was younger, another hypothesis that is difficult to
justify. \citet{Wick2000} and \citet{Ferrario2015a} have shown that
the field strength is independent of effective temperature as expected
by the very long ohmic decay time scales of white dwarfs. The
cumulative distribution function of the effective temperatures of the
sample of HFMWDs of \citet[][see their Figure\,5]{Ferrario2015a}
appears to be smooth over the full range of effective temperatures
($4\,000\le T_{\rm eff}/{\rm K}\le 45\,000$\,K) suggesting that the
birthrate of HMWDs has not altered over the age of the Galactic
disc. However, the sample of HFMWDs at our disposition is neither
volume nor magnitude-limited and biases easily come into play.

Thus, should a future enlarged and less biased sample of HFMWDs
confirm that the incidence of magnetism among cool white dwarfs is
indeed substantially higher than among hot white dwarfs, then the
possibility of field generation by accretion of giant gaseous planets
on to an originally non-magnetic white dwarf may provide a solution to
this puzzle. \citet{nordhaus2011} found that discs formed from
  tidally disrupted companions with masses in the range $0.1-500$
  Jupiter masses can explain the presence of high fields in white
  dwarfs. Thus, the central issue is, once again, how the
  magnetic field can diffuse into the core of a white dwarf over an
  appropriate timescale. This is a key question that still
  needs to be quantitatively answered.

The other question concerns the likelihood for an old and
  presumably stable planetary system to be sufficiently perturbed to
  send planets inward to plunge into the white dwarf.
\citet{Farihi2011} have shown that the number of close stellar
encounters that can have an appreciable effect on the outer regions of
a planetary system by sending objects into highly eccentric orbits is
around 0.5\,Gyr$^{-1}$. That is, the probability is about 50 per cent
every 0.5\,Gyr$^{-1}$. Considering typical cooling times between 1.5
and 9\,Gyr, these close encounters become likely during the life of a
white dwarf. If this hypothesis is correct, we should expect all white
dwarfs hosting a large gaseous planet to develop a magnetic field at
some point in their lifetime.

\section{Conclusions}\label{Conc}

In paper\,I we discussed the evolution of HFMWDs resulting from two
stellar cores (one of which is degenerate) that merge during a phase
of common envelope evolution.  We fitted the observed mass
distribution of the HFMWDs and the incidence of magnetism among
Galactic field white dwarfs and found that the HFMWDs are well
reproduced by the merging hypothesis for the origin of magnetic fields
if $0.1\le\alpha\le 0.3$.  However in paper\,I we did not propose a
prescription that would allow us to assign a magnetic field strength
to each white dwarf. This task has been carried out and the results
presented in this paper. We have assumed that the magnetic field
attained by the core of the single coalesced star emerging from common
envelope evolution is proportional to the orbital angular velocity of
the binary at the point the envelope is ejected. The break-up
angular velocity is the maximum that can be achieved by a
compact core during a merging process and this can only be reached if
the merging stars are in a very compact binary, such as a merging DD system.

In our model there are two parameters that must be empirically
estimated. These are $B_0$, that is linked to the efficiency with
which the poloidal field is regenerated by the decaying toroidal field
(see Wickramasinghe et al. 2014) and the common efficiency
parameter $\alpha$.  A K--S test was carried out on the CDFs of the
observed and theoretical field distributions for a wide range of $B_0$
and $\alpha$ and we found that the observed field distribution is best
represented by models characterised by $B_0=1.35 \times10^{10}$\,G and
$\alpha=0.2$. Population synthesis studies of MCVs that make use of
the results obtained in this paper and paper\,I is forthcoming and we
shall show that the same $B_0$ can also explain observations of
magnetic binaries.

We have also speculated that close stellar encounters can send a giant
gaseous planet from the outer regions of a white dwarf's planetary
system into a highly eccentric orbit. The plunging of this
super-Jupiter into the white dwarf can generate a magnetic
field and thus provide an answer to why magnetism among cool white
dwarfs, and particularly among cool DZ\, white dwarfs, is higher than
among hot white dwarfs.

\section*{Acknowledgements}

GPB gratefully acknowledges receipt of an Australian Postgraduate
Award.  CAT thanks the Australian National University for supporting a
visit as a Research Visitor of its Mathematical Sciences Institute,
Monash University for support as a Kevin Watford distinguished visitor
and Churchill College for his fellowship.\\

\end{document}